\documentclass[aps,pra,graphicx,9pt,twocolumn,superscriptaddress]{revtex4-2}
\usepackage{amsmath,amscd,graphicx,graphicx,cases}
\usepackage{multirow,float,subfigure,appendix}
\usepackage{diagbox,makecell,blindtext,color,amssymb}
\usepackage[colorlinks,linkcolor=blue,urlcolor=blue,
anchorcolor=blue,citecolor=blue]{hyperref}
\usepackage{textcomp,booktabs,colortbl,easyReview}
\usepackage{times}
\usepackage{xcolor,ulem,bm}
\usepackage{dcolumn}
\usepackage{mathrsfs}
\usepackage{amsfonts,multirow}
\usepackage{soul,lineno}
\usepackage{epstopdf}
\usepackage{verbatim}

\newif\ifistoreview

\begin{document}
	\title{An exceptional surface and its topology}
	\author{Shou-Bang Yang}
            \affiliation{Fujian Key Laboratory of Quantum Information and Quantum
        		Optics, College of Physics and Information Engineering, Fuzhou University,
        		Fuzhou, Fujian, 350108, China}
        \author{Pei-Rong Han} 
            \affiliation{School of Physics and Mechanical and Electrical Engineering, Longyan University, Longyan, 364012, China.}
	\author{Wen Ning}
            \affiliation{Fujian Key Laboratory of Quantum Information and Quantum
        		Optics, College of Physics and Information Engineering, Fuzhou University,
        		Fuzhou, Fujian, 350108, China}
	\author{Fan Wu} \email{t21060@fzu.edu.cn}
            \affiliation{Fujian Key Laboratory of Quantum Information and Quantum
		Optics, College of Physics and Information Engineering, Fuzhou University,
		Fuzhou, Fujian, 350108, China}
	\author{Zhen-Biao Yang} \email{zbyang@fzu.edu.cn}
            \affiliation{Fujian Key Laboratory of Quantum Information and Quantum
		Optics, College of Physics and Information Engineering, Fuzhou University,
		Fuzhou, Fujian, 350108, China}
	\author{Shi-Biao Zheng} 
            \affiliation{Fujian Key Laboratory of Quantum Information and Quantum
		Optics, College of Physics and Information Engineering, Fuzhou University,
		Fuzhou, Fujian, 350108, China}
	
	\vskip0.5cm
	
	\narrowtext
	
	\begin{abstract}
    Non-Hermitian (NH) systems can display exceptional topological defects without Hermitian counterparts, exemplified by exceptional rings in NH two-dimensional systems. However, exceptional topological features associated with higher-dimensional topological defects have only recently come into attention. We here investigate the topology for the singularities in an NH three-dimensional system. We find that the third-order singularities in the parameter space form an exceptional surface (ES), on which all the three eigenstates and eigenenergies coalesce. Such an ES corresponds to a two-dimensional extension of a point-like synthetic tensor monopole. We quantify its topology with the Dixmier-Douady invariant, which measures the quantized flux associated with the synthetic tensor field.  We further propose an experimentally feasible scheme for engineering such an NH model. Our results pave the way for investigations of exceptional topology associated with topological defects with more than one dimension.
	\end{abstract}
    \maketitle
\section{Introduction}	
 Although most quantum-mechanical phenomena are observed by isolating the quantum systems from their surrounding environment so as to minimize the decoherence effects, some of which, for instance, those caused by the non-Hermitian (NH) effects, are closely related to intriguing features that are inaccessible in the Hermitian cases \cite{miri_exceptional_2019,ozdemir_paritytime_2019,bergholtz_exceptional_2021,ding_non-hermitian_2022}, and can be harnessed to improve the sensitivities of sensors \cite{chen_exceptional_2017,hodaei_enhanced_2017,PhysRevLett.131.260201}. The rich physics of NH systems is closely associated with exceptional points (EPs), which feature the coalescence of both the eigenenergies and eigenstates \cite{ozdemir_paritytime_2019,bergholtz_exceptional_2021,ding_non-hermitian_2022}. This enables EPs to display distinct properties compared to degeneracies of Hermitian systems, where the eigenenergies coalesce but the eigenstates can remain orthogonal. These unique NH features include spectral real-to-complex transitions \cite{PhysRevLett.86.787,PhysRevLett.104.153601,gao_observation_2015,zhang_observation_2017}, chiral behaviors \cite{PhysRevX.8.021066,doppler_dynamically_2016,xu_topological_2016,yoon_time-asymmetric_2018,PhysRevLett.126.170506,PhysRevLett.124.070402,ren_chiral_2022}, exceptional entanglement transitions \cite{PhysRevLett.131.260201} and NH topology \cite{bergholtz_exceptional_2021,ding_non-hermitian_2022}. The topology of the EPs can be characterized by the topological invariants, such as the Chern number \cite{PhysRevLett.118.045701}, the winding number \cite{han2024measuring}, etc, which have been measured in different classical systems \cite{zhou_observation_2018,tang_exceptional_2020,su_direct_2021,PhysRevLett.127.034301} and quantum systems \cite{PhysRevLett.127.090501,han2024measuring}.
	
	The topological features associated with non-Hermiticity are further enriched by the discovery of the extention of EPs, such as exceptional rings (ERs) \cite{PhysRevLett.118.045701,PhysRevB.99.121101,PhysRevLett.127.196801,PhysRevB.104.L161117,PhysRevResearch.2.043268,PhysRevB.100.245205,zhen_spawning_2015,cerjan_experimental_2019,PhysRevLett.129.084301,tang_realization_2023} and exceptional surfaces (ESs) \cite{PhysRevLett.123.237202,zhou_exceptional_2019}. When the certain control parameter of the Hamiltonian is extended from the real to the complex domain, the exceptional structures are changed, for instance, a single EP merges into an ER \cite{zhang2024observationtopologicaltransitionsassociated}. Such an ER can be considered as a synthetic ring-like Dirac monopole in the parametric space, where the associated topology can be characterized by the first Chern number obtained by integrating the Berry curvature over a closed two-dimensional surface encircling
	the ring, as well as by a quantized Berry phase associated with the integral of the Berry connection along a two-dimensional loop encircling the ring \cite{PhysRevLett.118.045701}.
	
	To date, ERs have been observed in both classical \cite{zhen_spawning_2015,cerjan_experimental_2019,PhysRevLett.129.084301,tang_realization_2023} and quantum systems \cite{zhang2024observationtopologicaltransitionsassociated}, but restricted to rings formed by EP2s. Yet, higher-order exceptional structures and their topological properties  have only recently come into attention \cite{PhysRevResearch.7.L012021,PhysRevLett.127.186602,PhysRevLett.127.186601,PhysRevResearch.6.023205}. Third-order singularities can be considered synthetic tensor monopoles \cite{PhysRevD.31.1921,orland_instantons_1982}, which are related to tensor gauge fields \cite{PhysRevD.9.2273,henneaux_p-form_1986}, as opposed to the Dirac monopoles that are associated with vector gauge fields. 
	The tensor gauge fields, among which one paradigm is the Kalb-Ramond gauge field \cite{PhysRevD.9.2273,henneaux_p-form_1986}, are of great importance not only for string theory \cite{PhysRevD.9.2273,henneaux_p-form_1986}, but also for topological field theories which are essential for topological insulators and superconductors \cite{hansson_superconductors_2004,cho_topological_2011,PhysRevB.93.155122}. Exotic properties of synthetic tensor monopoles have been investigated both theoretically \cite{PhysRevLett.121.170401,PhysRevB.99.045154,Weisbrich2021tensormonopolesin} and experimentally \cite{chen_synthetic_2022,PhysRevLett.126.017702}, but limited to the Hermitian systems.  
    However, NH counterparts of such topological defects remained unexplored so far \cite{doi:10.1126/sciadv.aat0346}. 
	
	In this work, we study the geometric features of the exceptional surface (ES), formed by EP3s of an NH three-dimensional system. Such an ES corresponds to the two-dimensional extension of a point-like tensor monopole in the four-dimensional parameter space of the NH Hamiltonian.
 The topological properties of the ES are characterized by the Dixmier-Douady (DD) invariant \cite{bouwknegt_d-branes_2000,PhysRevD.31.1921,orland_instantons_1982,chen_synthetic_2022,PhysRevLett.126.017702} as well as by the Berry phase, uncovering the exceptional Third-order topology with an ES structured in the physically controllable NH three-dimensional system. Our work provides an effective method for the characterization of the exceptional topology of two-dimensional topological defects, paving the way for deep exploration into higher-order topological physics in NH systems.
	
	

	\begin{figure*}[t!] 
		\centering
		\includegraphics[width=6.8in]{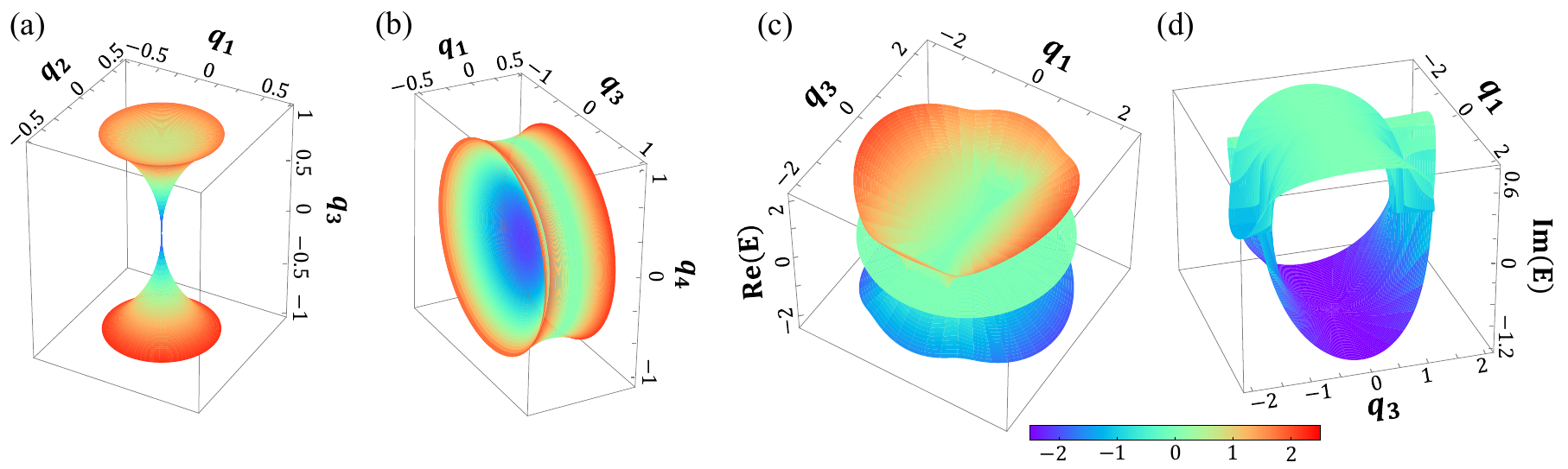}
		\caption{The projection of the ES from the four-dimensional space onto the three-dimensional case, which is defined by the coordinates $\{q_1,q_2,q_3 \}$ $(q_4=0)$ in (a) and $\{q_1,q_3,q_4 \}$ $(q_2=0)$ in (b). Real (c) and imaginary (d) parts of the eigenenergies with respect to $\Omega_1=q_1+iq_2$ and $\Omega_2=q_3+iq_4$.}
		\label{Fig1}
	\end{figure*}%
\section{Structure of the ES}
A singularity in the three-dimensional Hermitian system can extend into an ES provided an NH dynamics is involved. The NH Hamiltonian for such a three-dimensional system can be written as 
	\begin{align}
		H/\hbar&= \vec{q}\cdot \vec{\lambda}+i\kappa\lambda_8,
	\label{eq1}
	\end{align}
	where $\vec{q}=\{q_1,q_2,q_3,q_4\}$ determines the four-dimensional parameter space and $\vec{\lambda}=\{\lambda_1,\lambda_2,\lambda_6,\lambda_7\}$ are the $3\times3$ Gell-Mann matrices \cite{PhysRev.125.1067}, which satisfy the relation $[\lambda_j,\lambda_k]=if^{jkl}\lambda_l$ (see Supplemental Material).

	The complex eigenenergies of the Hamiltonian (\ref{eq1}) are
	\begin{small}
	\begin{eqnarray}
		E_1&=&-\frac{\sqrt{3}}{3}i\kappa\left[\sqrt[3]{(A+C)/2}+\frac{B}{\sqrt[3]{(A+C)/2}}\right],\nonumber\\
		E_2&=&\frac{\sqrt{3}}{3}i\kappa\left[\frac{(1-\sqrt{3}i)\sqrt[3]{(A+C)/2}}{2}+
		\frac{(1+\sqrt{3}i)B}{2\sqrt[3]{(A+C)/2}}\right],\nonumber\\
		E_3&=&\frac{\sqrt{3}}{3}i\kappa\left[\frac{(1+\sqrt{3}i)\sqrt[3]{(A+C)/2}}{2}+
		\frac{(1-\sqrt{3}i)B}{2\sqrt[3]{(A+C)/2}}\right],\nonumber\\
	\end{eqnarray}
	\end{small}
where 
\begin{eqnarray}
	A&=&6|\Omega_1|^2-3|\Omega_2|^2+2\kappa^2,\nonumber\\ 
	B&=&|\Omega_1|^2+|\Omega_2|^2-\kappa^2,\nonumber\\
	C&=&\sqrt{4B^3-A^2},
\end{eqnarray}
with $\Omega_{1}=q_1+iq_2$ and $\Omega_{2}=q_3+iq_4$ being the coupling strengths between neighboring states.

For $\kappa=0$, the Hamiltonian (\ref{eq1}) is Hermitian and has three eigenvectors corresponding to three real eigenenergies, $E_n=0,\pm\sqrt{|\Omega_1|^2+|\Omega_2|^2}$, with a singularity located at  $\vec{q}=\{0,0,0,0\}$. Such a three-fold degeneracy is referred to as the tensor monopole in the parameter space \cite{PhysRevD.9.2273,henneaux_p-form_1986}. When the NH term of $i\kappa{\lambda_8}$ is introduced, the eigenenergies of the system become complex, while the eigenvectors are not orthogonal. The point-like tensor monopole morphs into a three-order ES located in the four-dimensional parameter space of $\{q_1,q_2,q_3,q_4\}$, with $|\Omega_1|=\kappa/3$ and $|\Omega_2|=2\sqrt{2}\kappa/3$. Furthermore, the ES tends to be closed, which is double-degenerate in the case of $E_i=E_j$ when $B=0$. The projections of this ES onto the three-dimensional spaces of $\{q_1,q_2,q_3\}$ $(q_4=0)$ and $\{q_1,q_3,q_4\}$ $(q_2=0)$ are given in Fig.~\ref{Fig1}(a) and Fig.~\ref{Fig1}(b), respectively, with the assumption of $\kappa=1$.
	
	The closed ES is filled by a four-dimensional bulk Fermi arc, along which the real parts of the three complex eigenenergies degenerate to the value of 0. For ${q_2}={q_4}=0$,
	the real and imaginary parts of the complex eigenenergies are shown in Fig.~\ref{Fig1}(c) and Fig.~\ref{Fig1}(d). Outside the Fermi arc, the real parts of the eigenenergies extend gradually from zero without overlapping each other, while the imaginary parts remain fixed. In contrast, inside the Fermi arc, the real parts of the eigenenergies converge to zero, while the imaginary ones differ from each other. The exceptional features exhibited by such distinct structures, i.e., from an open ES to a closed ES, essentially reflect the peculiar symmetries in a three-dimensional NH system.   
	
 \section{The DD invariant}
 It is intriguing to explore the topology inherent in such a non-trivial symmetry. There exist abundant topological geometries in the energy band structure of the system, such as the ES, the hyperboloids composed of EP2s connecting the ES, and the Fermi arc regions containing them, etc. The ES is formed in the parameter space due to the introduction of the NH term, which results in a tensor node in the four-dimensional parameter space. Similar to the topological defect associated with a nodal point for the Hermitian case in the equal dimension, the DD invariant for a third-order ES involves the flux of a radial three-form curvature tensor $M_{\mu\nu\lambda}$ over a four-sphere that surrounds this ES, which is the generalization of the two-form Berry curvature $F_{\mu\nu}$ of the Dirac monopole, i.e., 
	\begin{equation}
		\mathcal{DD}=\frac{1}{2\pi^2}\int_{S^3}M_{\mu\nu\lambda}\textrm{d}q^\mu\wedge \textrm{d}q^\nu\wedge \textrm{d}q^\lambda.
        \label{eq4}
	\end{equation}
	In (\ref{eq4}), the three-form curvature tensor, ${M}_{\mu v \lambda}$, is related to the quantum metric or the two-form curvature as $F_{jk}$ $(j, k =\mu, v, \lambda)$ \cite{PhysRevLett.121.170401,PhysRevB.99.045154}, and described as
	\begin{equation}
		M_{\mu\nu\lambda}=\epsilon_{\mu\nu\lambda}\left[4\sqrt{\textrm{det}(g_{\mu\nu\lambda})}\right]
		=-\frac{1}{2}\left(\mathcal{F}_{\mu\nu}+\mathcal{F}_{\lambda\mu}\right).
		\label{eq04}
	\end{equation}
	\begin{figure}[t!] 
		\centering
		\includegraphics[width=3.2in]{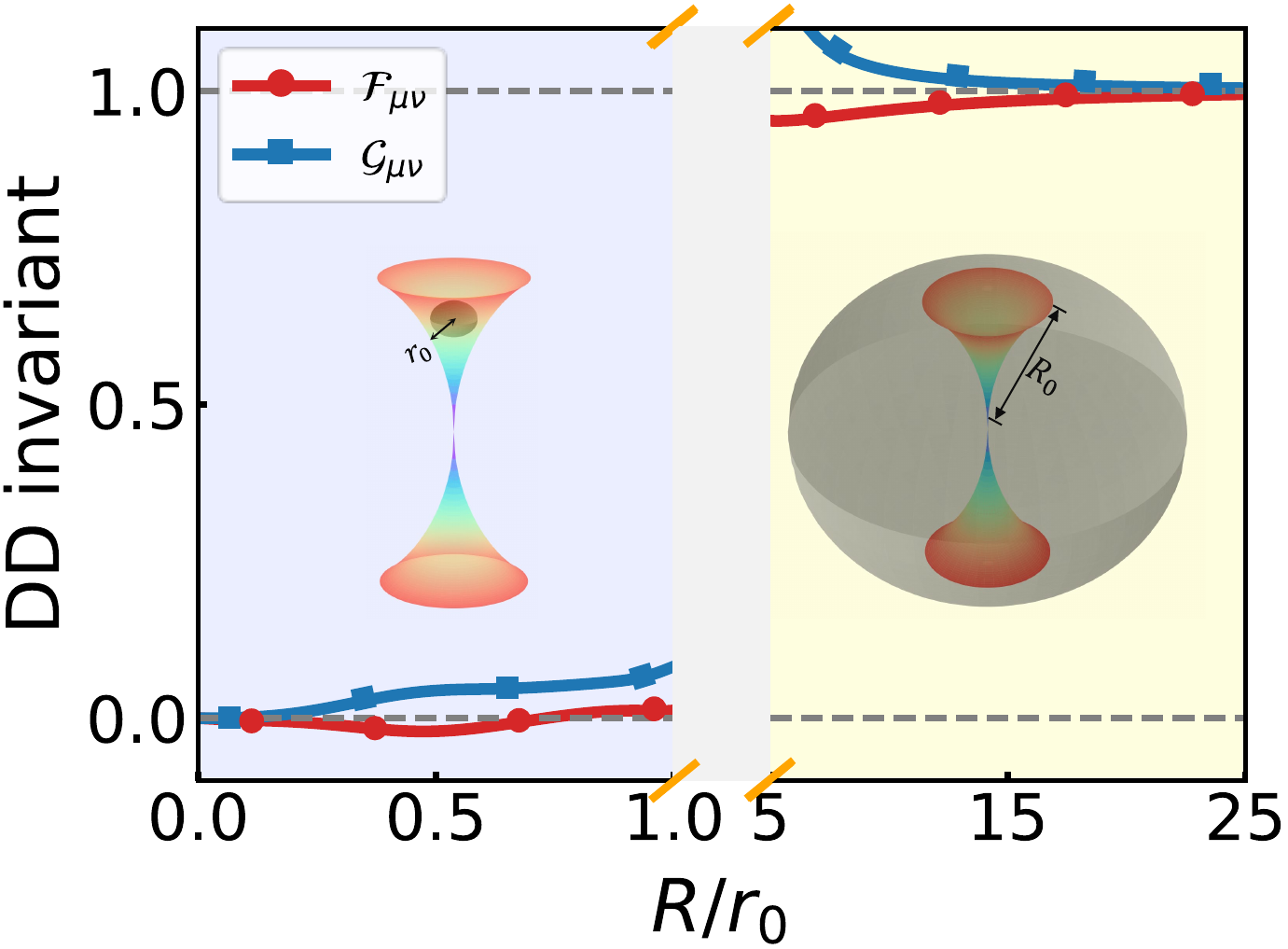}
		\caption{The DD invariant characterized in the NH three-dimensional system, with respect to $R/r_0$, where $R$ is the radius of the parameter sphere and $r_0$ is the shortest distance between the sphere's center and the ES. The left and right sides of the broken shaft indicate the two cases of the parameter sphere being inside and outside the ES as the radius $R$ increases, respectively. The blue and red lines characterize the DD invariant calculated by the quantum metric and the Berry curvature described in Eq. (\ref{eq04}), respectively. The grey dashed lines are the theoretical horizontal lines of the values 0 and 1.}
		\label{Fig3}
	\end{figure}%
	
	The quantum metric tensor and the Berry curvature respectively correspond to the real and imaginary parts of the quantum geometric tensor, which is written as (see Supplemental Material)
	\begin{eqnarray}
		\chi_{\mu\nu}=g_{\mu\nu}+i\mathcal{F}_{\mu\nu}=\sum_{n\neq-}\frac{|\langle \psi_{-}^{L}|\partial_\mu H|\psi_n\rangle\langle \psi_{-}^{L}|\partial_\nu H|\psi_n\rangle|}{\left(E^L_{-}-E_n\right)\left(E_{-}-E_n^L\right)},\nonumber\\
		\label{eq8}
	\end{eqnarray}
	where $\langle\psi_n^L|$ denotes the normalized left eigenvector of $|\psi_n\rangle$ and satisfies $\langle\psi_n^L|\psi_m\rangle=\delta_{mn}$, and $E_n^L$ is the eigenenergy of $\langle\psi_n^L|$ satisfying $\langle\psi_n^L|H=\langle\psi_n^L|E_n^L$.
	
	We set $\Omega_{1,2}$ in a physically controllable way as  $\{\Omega_1=R\textrm{cos}(\alpha)e^{i\beta}, \Omega_2=R\textrm{sin}(\alpha)e^{i\phi}\}$  to construct the NH Hamiltonian in the four-dimensional parameter space, where $R$ is the radius of the four-dimensional parameter sphere and $r_0$ represents the shortest distance between the sphere's center and the ES. For $R<r_0$, the parameter sphere is located inside the ES; while for $R>R_0$, the parameter sphere wraps around the whole ES. When $R<r_0$, the parameter sphere is enclosed by the ES which is determined by three parameters $\alpha$, $\beta$ and $\phi$. The schematic representation of the four-dimensional parameter space manifold projected onto the three-dimensional one is shown in Fig.~\ref{Fig3}. The gray sphere denotes the projection structure of the four-dimensional parameter space onto the three-dimensional case of $\{q_1,q_2,q_3 \}$, and the colored closed surface shows the ES as in Fig.~\ref{Fig1}(a). 
	
	The DD invariant obtained from the three-form curvature is given by Fig.~\ref{Fig3}, where the red line shows the result of the two-form Berry curvature while the blue line depicts the result of the quantum metric. When the three-dimensional parameter manifold, which is of one order reduced, is outside of the Fermi arc and wraps around the whole ES, the DD invariant is calculated as $\mathcal{DD} = 1$, indicating the system is topologically non-trivial; when the parameter manifold gradually shrinks and eventually falls inside the whole Fermi arc, the DD invariant vanishes and the system reduces to a trivial one. This result indicates the exceptional topology of the ES in such a three-dimensional NH system, as compared to the three-dimensional Hermitian case. The discontinuity of the DD invariant in the broken axis in Fig.~\ref{Fig3} is precisely due to the introduction of the NH term, which transforms a singularity into an ES, resulting in a peculiar symmetry. Due to the peculiar structure of the ES and chosen 3-sphere manifold, it shows that there approximately exists a region of $R/r_0 \in [1,5]$ where such a topology is ill defined, leading to the ambiguous DD invariant therein. The DD invariant only approaches the theoretical expectation when it gets far away from the region.
	
 \section{The Berry phase}
 In addition to the DD invariant, the Berry phase can also be used to characterize the topology of the established ES, which can be projected onto a two-dimensional ER.
	\begin{figure}[b!] 
		\centering
		\includegraphics[width=3.2in]{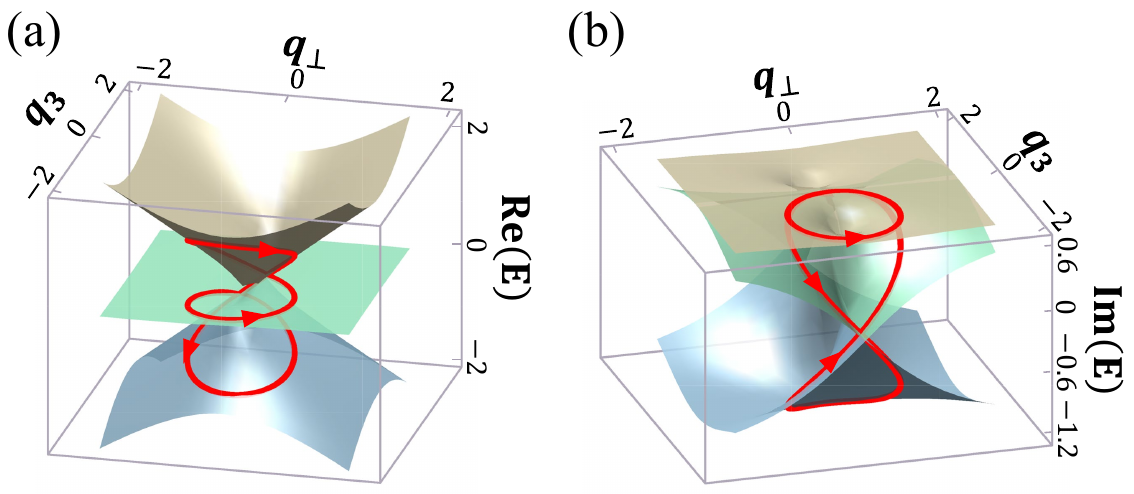}
		\caption{Energy spectra and the Riemann surface of the NH model (\ref{eq1}). The real (a) and imaginary (b) parts of the Riemann surface with respect to $q_3$ and $q_{\perp}$. The red arrowhead tube represents the travel path from $\theta=0$ to $\theta=6\pi$.}
		\label{Fig4}
	\end{figure}%
	In contrast to the Hermitian system, as the NH system possesses complex eigenenergies and peculiar eigenvectors, the involved Berry phase exhibits unique characteristics. Such a Berry phase is defined as
	\begin{eqnarray}
		\mathcal{P}=\oint_{3\mathcal{L}}i\langle\psi_n^L|\partial_\theta\psi_n\rangle\textrm{d}\theta,
		\label{eq10}
	\end{eqnarray}
	where $\langle\psi^L_i|$ and $|\psi^R_i\rangle$ are the left and right eigenvectors,
and the path $3\mathcal{L}$ encircles the ring three times along the Riemann surface so that the eigenvector travels and finally returns to the origin. For the four-dimensional ES projected on the two-dimensional parameter space of $\{\vec{q_3},\vec{q_4}\}$, we construct an effective control to enable the production of the Berry phase. The control Hamiltonian is modeled as
	\begin{eqnarray}
		H_{\mathcal{P}}/\hbar = \frac{\kappa}{3}\lambda_1+q_3\lambda_6+q_{\perp}\left(\lambda_8-\frac{1}{\sqrt{3}}I\right)+i\kappa\lambda_8,
		\label{eq11}
	\end{eqnarray}
	where $\lambda_n$ is the nth Gell-Mann matrix \cite{PhysRev.125.1067} and $\vec{q_{\perp}}$ is along the axis perpendicular to the $\{\vec{q_3},\vec{q_4}\}$ plane. The evolution path can be guaranteed provided the condition of ($q_3= R\textrm{sin}\theta+\Delta$, $q_4=0$, $q_{\perp}=R\textrm{cos}\theta$) with $(\Delta=2\sqrt{2}/3, R=0.85)$ is met. Notice here that a shift of $\Delta$ is added in $\vec{q_3}$ for the convenience of physical control. The entire trajectories of the $3\mathcal{L}$ path in real and imaginary parts of the Riemann surface are shown in Fig.~\ref{Fig4}(a) and Fig.~\ref{Fig4}(b), respectively. Each of the eigenenergies circles along the parameter loop three times and then returns to the original value, forming a peculiar three-sided Möbius-like structure (see Supplemental Material).
	 \begin{figure}[t] 
	 	\centering
	 	\includegraphics[width=3.0in]{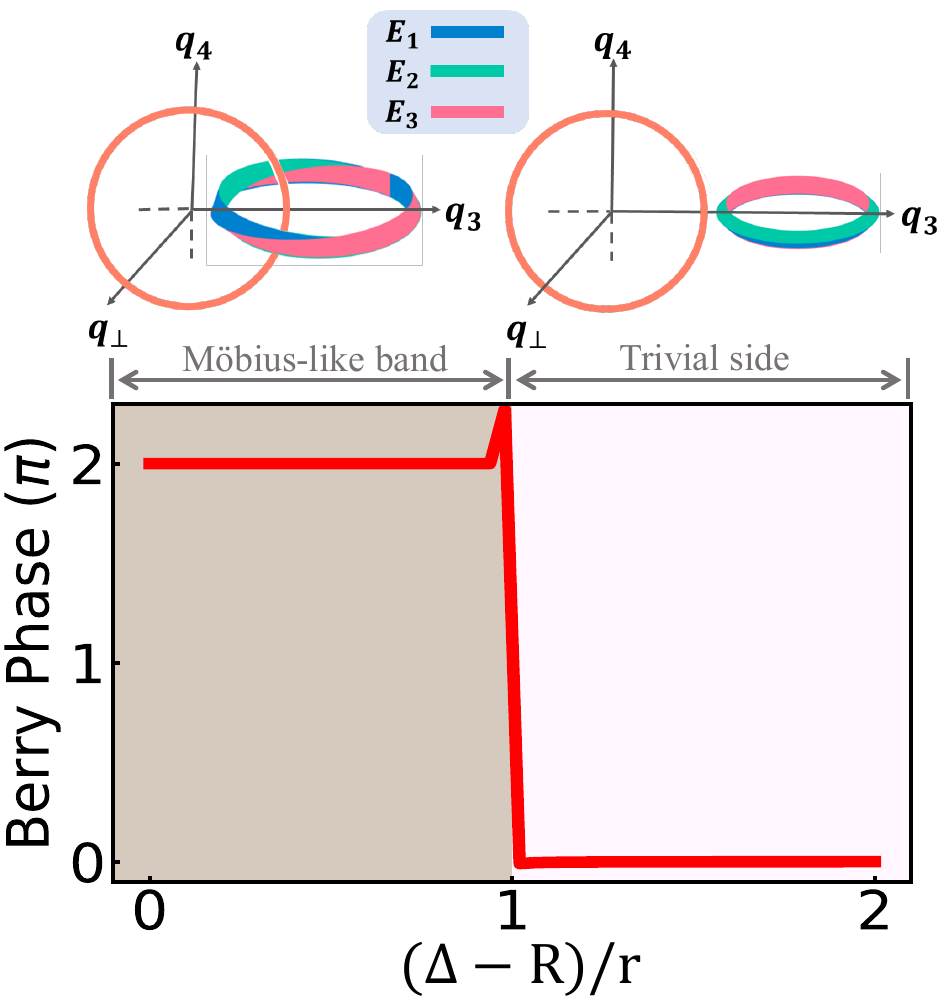}
	 	\caption{The Berry phase characterized in the NH model. The parameter loop on the $\{\vec{q_3},\vec{q_\perp}\}$ plane travels encircling (upper-left) or separated from (upper-right) the ER on the $\{\vec{q_3},\vec{q_4}\}$ plane. The Berry phase versus $(\Delta-R)/r$, a sharp transition happens when the parameter loop shrinks to pass through the ER.}
	 	\label{Fig5}
	 \end{figure}%
	
	As the parameter gradually travels along the $3\mathcal{L}$ path surrounding the ER, the Berry phase accumulated yields $\mathcal{P}=2\pi$ ultimately, as shown in the left panel of Fig.~\ref{Fig5}. The diagram on the top left panel shows the law of the three-sided Möbius-like eigenenergies, exhibited by encircling the ER according to the parameter change. Such a Berry phase is the extention of two-sided M$\ddot{o}$bius-like eigenenergies of the ER in the two-dimensional parameter space to a higher-dimensional case, both different from the case with the nodal ring \cite{PhysRevB.84.235126,deng_nodal_2019}. On the other hand, $\Delta$ can be slowly changed so that the parameter loop no longer passes through the ER. The diagram on the top right panel shows that the Möbius-like structures of the eigenenergies disappear and the Berry phase turns to 0. At the critical point when $\Delta-R=r$ with $r=2\sqrt{2}/3$ being the radius of ER, a topological transition happens where the Berry phase jumps from $2\pi$ to 0.
	
	\section{Bulk-boundary correspondence}
    \begin{figure}[b!] 
	 	\centering
	 	\includegraphics[width=3.2in]{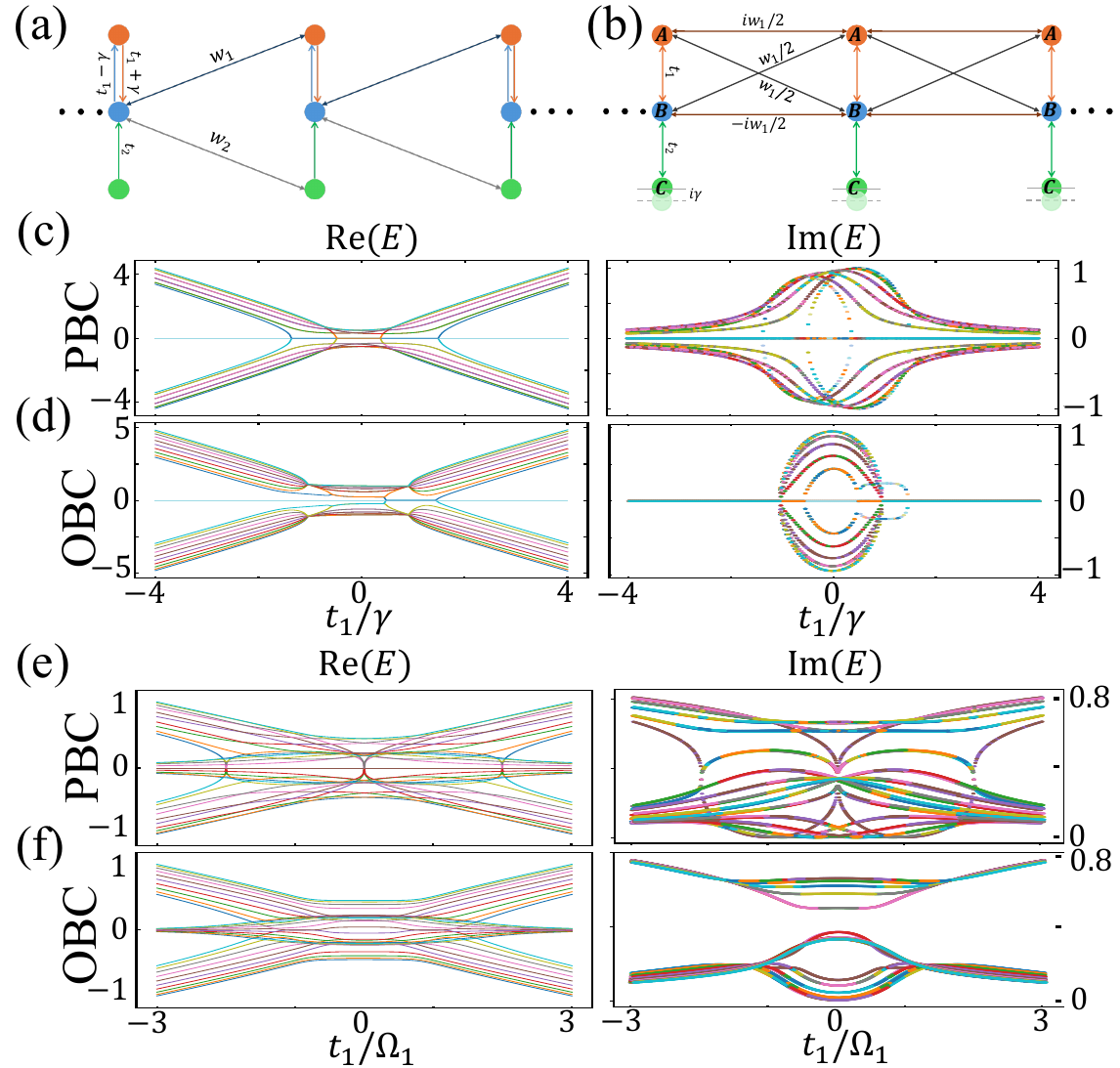}
	 	\caption{Topological boundary modes in the trimer SSH model. (a) and (b) are two distinct toy models of the NH SSH3 system, each defined by a unique bulk Hamiltonian. For the first model, the energy spectra are under PBC (c) and OBC (d), respectively. Under OBC, the NHSE is clearly observed. In the second model, the EP3s are observed under PBC (e), whereas they vanish under OBC (f).}
	 	\label{Fig6}
	 \end{figure}%
    The investigation of the bulk-boundary correspondence constitutes a central issue in the study of non-Hermitian topological states \cite{PhysRevLett.121.086803,PhysRevLett.116.133903,xiong_why_2018}. We consider a NH trimer Su-Schrieffer-Heeger (SSH3) model, schematically illustrated in Fig.~\ref{Fig6}(a). Under periodic boundary conditions (PBC), its bulk Hamiltonian is denoted as 
    \begin{eqnarray} H_k&=&\left(t_1+w_1\cos{k}\right)\lambda_1+\left(w_1\sin{k}+i\gamma\right)\lambda_2\nonumber\\
    &&+\left(t_2+w_2\cos{k}\right)\lambda_6+\left(w_2\sin{k}\right)\lambda_7,
    \label{eqn9}
    \end{eqnarray}
    where $t_{1,2}$, $\gamma$ denote the hopping between the sites in a unit cell, $w_{1,2}$ are the coupling between different unit cells and $k=2\pi n/N$ for each unit cell.
    The energy gap closes at exceptional points when $\left(t_1+w_1\right)^2+\left(t_2+w_2\right)^2=\gamma^2$ $\left(k=0\right)$ or $\left(t_1-w_1\right)^2+\left(t_2-w_2\right)^2=\gamma^2$ $\left(k=\pi\right)$. For a system size of $N = 10$ unit cells and fixed parameters $t_2=\gamma/4$, $w_1=\gamma$ and $w_2=\gamma/4$, the distribution of the real and imaginary parts of the energy spectra are shown in Fig.~\ref{Fig6}(c), resembling that of the NH SSH model under PBC \cite{PhysRevLett.116.133903}. The relevant topological invariant can also be characterized by a winding number. Under open boundary conditions (OBC), the distributions of the real and imaginary parts of the energy spectrum are displayed in Fig.~\ref{Fig6}(d). Consistent with the lower-dimensional behavior \cite{PhysRevLett.121.086803}, a large number of eigenstates in the open chain are found to be spatially localized near the boundaries, indicative of the NH skin effect (NHSE). In this regime, conventional Bloch topological invariants become inadequate, necessitating the introduction of non-Bloch topological invariants, such as the non-Bloch winding number \cite{PhysRevLett.121.086803}, to properly determine the topological boundary modes.
    
    To further develop the model, we modify the aforementioned SSH3 model by introducing the NH term on the diagonal element of the bulk Hamiltonian, this modification yields a bulk Hamiltonian which closely aligns with the model discussed in Eq.~(\ref{eq1}), thereby enabling the simulation of the topological defect associated with the ES, with the toy model shown in Fig.~\ref{Fig6}(b). Under the PBC, the bulk Hamiltonian is given by
    \begin{small}
    \begin{eqnarray} H_k^\prime&=&\left(t_1+w_1\cos{k}\right)\lambda_1+\left(w_1\sin{k}\right)\lambda_3\nonumber\\
    &&+t_2\lambda_6-i\gamma\left(\lambda_8-I/\sqrt{3}\right).
    \label{eqn10}
    \end{eqnarray}
    \end{small}
    Notably, in such a configuration of Eq.~(\ref{eqn10}), the energy gap closes at EP3s when $t_1+w_1=\pm\Omega_1$ $(k=0)$ or $t_1-w_1=\pm\Omega_1$ ($k=\pi$) under $t_2=\pm\Omega_2$, where $\Omega_1=\gamma/3\sqrt{3}$ and $\Omega_2=2\sqrt{2}\gamma/3\sqrt{3}$. Similarly, with the fixed parameters $t_2=\Omega_2$ and $w_1=\Omega_1$, the real and imaginary parts of the energy spectra in this SSH3 system as functions of $t_1$ ($\in[-3\Omega_1,3\Omega_1 ]$), as shown in Fig.~\ref{Fig6}(e). The topological properties of the system can be characterized by the winding number, i.e., $W=2/3$, corresponding to the Berry phase of $P_{3\mathcal{L}}=2\pi$. Under the OBC, the distribution of spectra (real and imaginary parts) are displayed in Fig.~\ref{Fig6}(f). Unlike the first case of Eq.~(\ref{eqn9}), where the NHSE is absent, and more notably, the EP3s also vanish.
    
\section{A proposal for experimental implementation}
 In order to characterize the exceptional topology of the ES experimentally, we may consider a model in which two resonators $R_1$ and $R_2$ are coupled to a qubit $Q$, the dynamics of the system comprising two resonators, qubit, resonator decay and dephasing, qubit decay and dephasing, can be modeled utilizing the Lindblad master equation
 \begin{small}
        \begin{eqnarray}
			\frac{d\rho(t)}{dt}&=&-\frac{i}{\hbar}[H(t),\rho(t)]  \nonumber \\
            && + \Gamma_d D[\hat{\sigma}^-]\rho(t) +\Gamma_p/2 D[\hat{\sigma}_z]\rho(t)  \nonumber \\
            && + \sum_{j=1,2}\left\{\kappa_{d,j} D[\hat{a}_j]\rho(t) + \kappa_{p,j}D[\hat{a}_j^{\dagger}\hat{a}_j]\rho(t)\right\}, 
        \label{eq99}
		\end{eqnarray}
\end{small}
 where the Hamiltonian is written as 
 \begin{small}
		\begin{eqnarray}
			H(t)/\hbar &=& \frac{\omega_q}{2}\hat{\sigma}_z+\sum_{j=1,2}\omega_{r_j}\hat{a}_j^{\dagger}\hat{a}_j+ \sum_{j=1,2} \left(g_j \hat{a}_j^\dagger \hat{\sigma}^- + H.c.\right).\nonumber\\
			\label{eq10}
		\end{eqnarray}
\end{small}
	Here, $\omega_{q}/{2\pi}$ and $\omega_{r_j}/{2\pi}$ represent the eigenfrequencies of $Q$ and $R_j$ $(j = 1, 2)$, respectively, $\hat{\sigma}_z$ is the qubit inversion operator defined as $\hat{\sigma}_z = \hat{\sigma}^{\dagger} \hat{\sigma}^{-} - \hat{\sigma}^{-} \hat{\sigma}^{\dagger}$, where $\hat{\sigma}^{\dagger}$ ($\hat{\sigma}^{-}$) being the qubit raising (lowing) operator, $\hat{a}_j^{\dagger}$ ($\hat{a}_j$) is the creation (annihilation) operator for the photon of resonator $j$, $g_j$ is the coupling strength between qubit and resonator $j$, $\Gamma_d$ ($\Gamma_p$) and $\kappa_{d,j}$ ($\kappa_{p,j}$) are the qubit energy relaxation (pure dephasing) rate and the resonator $j$ photon energy relaxation (pure dephasing) rate, respectively, and the Lindblad super-operator is defined as $L[A]\rho = A\rho A^{\dagger}-\frac{1}{2}A^{\dagger}A\rho - \frac{1}{2}\rho A^{\dagger}A$ for any dissipator operator $A$ ($A= \hat{\sigma}^-, \hat{\sigma}_z, \hat{a}_j, \hat{a}_j^{\dagger}\hat{a}_j$). Here the thermal excitations for the qubit and resonators are assumed to be negligible. 
 
    \begin{figure}[t!] 
		\centering
		\includegraphics[width=3.2in]{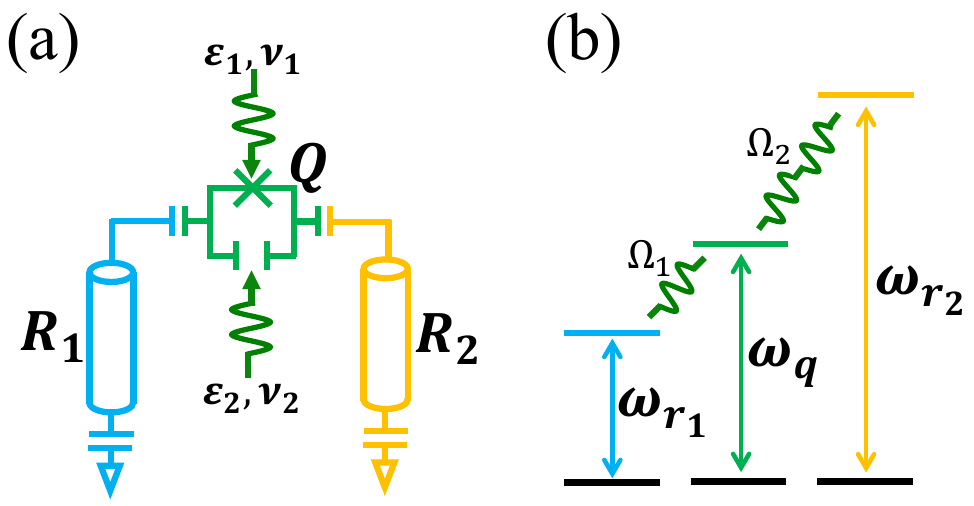}
		\caption{Implementation of the NH topological model. (a) Device schematics. A lossless resonator ($R_1$) and a lossy resonator ($R_2$), are coupled to a qubit ($Q$). The frequencies for $R_1$, $R_2$, and $Q$ are $\omega_1/2\pi$, $\omega_2/2\pi$, and $\omega_q/2\pi$, respectively. Two modulation pulses are applied to the qubit to control the coupling coefficients $\Omega_{1,2}$. (b) The modulation of pulses. The interaction between the qubit and two resonators is achieved by the control of amplitudes $\varepsilon_{1,2}$ and frequencies $\nu_{1,2}$ of the modulation pulses.}
		\label{Fig55}
	\end{figure}%

  We adopt the method realized by our previous experimental implementation \cite{PhysRevLett.131.260201, han2024measuring}, where the qubit (Q) is coupled to a lossless bus resonator ($R_1$) and its lossy readout resonator ($R_2$) \cite{song_continuous-variable_2017,PhysRevLett.123.060502,yang_experimental_2021}, as depicted in Fig.~\ref{Fig55}(a). In such two experiments \cite{PhysRevLett.131.260201, han2024measuring}, qubit energy relaxation and dephasing time, resonator $R_1$ energy relaxation and dephasing time, and resonator $R_2$ dephasing time, are much larger than the relevant time scales of the dynamical evolution, while resonator $R_2$ energy relaxation plays an equivalently dominant role as the unitary evolution for the whole dynamics. Under the condition that the system state evolution trajectory is subject to no photon-number jumps, the system dynamics governed by Eq. (\ref{eq99}) is reduced to
\begin{small}
\begin{equation}\label{eq11}
H^{\prime}(t)/\hbar = H(t)/\hbar-\frac{i}{2}\kappa_{d,2} \hat{a}_2^\dagger \hat{a}_2, 
\end{equation}
\end{small}
which in practice can be achieved through postselection \cite{PhysRevLett.131.260201, han2024measuring}.
  In order to realize the controllable parameter manifolds and loops, we apply two modulation pulses to $Q$ with the form of \cite{PhysRevLett.131.260201, han2024measuring}
  \begin{small}
	\begin{eqnarray}
		\omega_e(t)=\omega_q+\varepsilon_1\textrm{cos}(\nu_1 t)+\varepsilon_2\textrm{cos}(\nu_2 t),
		\label{eq15}
	\end{eqnarray}
    \end{small}
	where $\varepsilon_{1,2}$ and $\nu_{1,2}$ denote the modulation amplitudes and frequencies, respectively \cite{ning_experimental_2024}. 
	We assume that the frequencies of the three subsystems satisfy $\omega_{r_2}>\omega_{q}>\omega_{r_1}$, as shown in Fig.~\ref{Fig55}(b). In order to induce resonant coupling of $Q$ to both $R_1$ and $R_2$, the modulation frequencies are set as $\nu_1=\Delta_1/2$ and $\nu_2=-\Delta_2$, respectively, where $\Delta_j = \omega_q - \omega_{r_j} $. With the terms of fast oscillations being discarded, the conditional Hamiltonian (\ref{eq11}) in the interaction picture is reduced to $H_I^{\prime}/\hbar=[g_1J_2(\mu_1)\hat{a}_1^\dagger|g\rangle\langle e|+g_2J_{-1}(\mu_2)\hat{a}_2^\dagger|g\rangle\langle e|+H.c.]-\frac{i}{2}\kappa_{d,2} \hat{a}_2^\dagger \hat{a}_2$, where $J_2(\mu_1)$ and $J_{-1}(\mu_2)$ are the $2$th and $(-1)$th Bessel functions of the first kind, with $\mu_j = \varepsilon_n/\nu_n$ ($n = 1,2$) (see Supplemental Material). By adjusting $\varepsilon_{1,2}$ and $\nu_{1,2}$ to satisfy $g_1J_2(\mu_1)=\Omega_1$ and $g_1J_{-1}(\mu_2)=\Omega_2$, measuring the states of the system at different evolution times, and postselecting the bases of the single-excitation, belonging to the Hilbert subspace of $\{|e00\rangle,|g10\rangle,|g01\rangle\}$ \cite{han2024measuring}, we can fit out accordingly the right- and left-eigenvectors and the corresponding eigenenergies \cite{PhysRevLett.131.260201}, based on which the DD invariant and the Berry phase can be extracted.  

 \section{Conclusion}
	In summary, we have investigated the geometric features of the ES in the four-dimensional parameter space of an NH three-dimensional system. The topology is characterized by the DD invariant, as well
as by the Berry phase. The DD invariant is 1 when the parameter-phase manifold encloses the ES, but becomes 0 when the manifold is inside the ES. The Berry phase associated with a loop is either 2$\pi$ or 0, depending upon whether or not the loop encircles the ER, which corresponds to a one-dimensional projection of the ES. We have further proposed a protocol to experimentally realize the NH three-dimensional model in the superconducting circuit architecture, where the NH three-band system can be established by utilizing a frequency-tunable superconducting qubit modulably coupled to a lossless resonator and a lossy resonator, combined with the use of postselection on the system state confined within the subspace subjected to no quantum jump.     
    
    Our study provides an effective method for the characterization of the ES topology by resorting to the DD topological invariant in the four-dimensional parameter space and the Berry phase in the projected two-dimensional parameter space, stimulating the study of the exceptional higher-order topology of the NH systems. Recent development of the superconducting circuit cavity quantum electrodynamics techniques provides the potential for the experimental realization of the protocol \cite{PhysRevLett.131.260201,han2024measuring,zhang2024observationtopologicaltransitionsassociated}.

This work was supported by the National Natural Science Foundation of China (Grant Nos. 12474356, 12475015, 12274080, 12204105, 11875108).

    Conflict of Interest\textemdash The authors declare that they have no conflict of interest.
 
\bibliography{reference}
\end{document}

\bibliography{reference}
\end{document}


\title{Supplemental Material for "An exceptional surface and its topology"}
    	\author{Shou-Bang Yang}
            \affiliation{Fujian Key Laboratory of Quantum Information and Quantum
        		Optics, College of Physics and Information Engineering, Fuzhou University,
        		Fuzhou, Fujian, 350108, China}
        \author{Pei-Rong Han} 
            \affiliation{School of Physics and Mechanical and Electrical Engineering, Longyan University, Longyan, 364012, China.}
	\author{Wen Ning}
            \affiliation{Fujian Key Laboratory of Quantum Information and Quantum
        		Optics, College of Physics and Information Engineering, Fuzhou University,
        		Fuzhou, Fujian, 350108, China}
	\author{Fan Wu} \email{t21060@fzu.edu.cn}
            \affiliation{Fujian Key Laboratory of Quantum Information and Quantum
		Optics, College of Physics and Information Engineering, Fuzhou University,
		Fuzhou, Fujian, 350108, China}
	\author{Zhen-Biao Yang} \email{zbyang@fzu.edu.cn}
            \affiliation{Fujian Key Laboratory of Quantum Information and Quantum
		Optics, College of Physics and Information Engineering, Fuzhou University,
		Fuzhou, Fujian, 350108, China}
	\author{Shi-Biao Zheng} 
            \affiliation{Fujian Key Laboratory of Quantum Information and Quantum
		Optics, College of Physics and Information Engineering, Fuzhou University,
		Fuzhou, Fujian, 350108, China}
	
	\vskip0.5cm
    \maketitle
	\tableofcontents
\renewcommand{\thefigure}{S\arabic{figure}}
\setcounter{figure}{0}
\section{Topological properties in non-Hermitian (NH) systems}
	\subsection{Dixmier-Douady invariants in the ES}
 \begin{figure*}[t!] 
		\centering
		\includegraphics[width=6.2in]{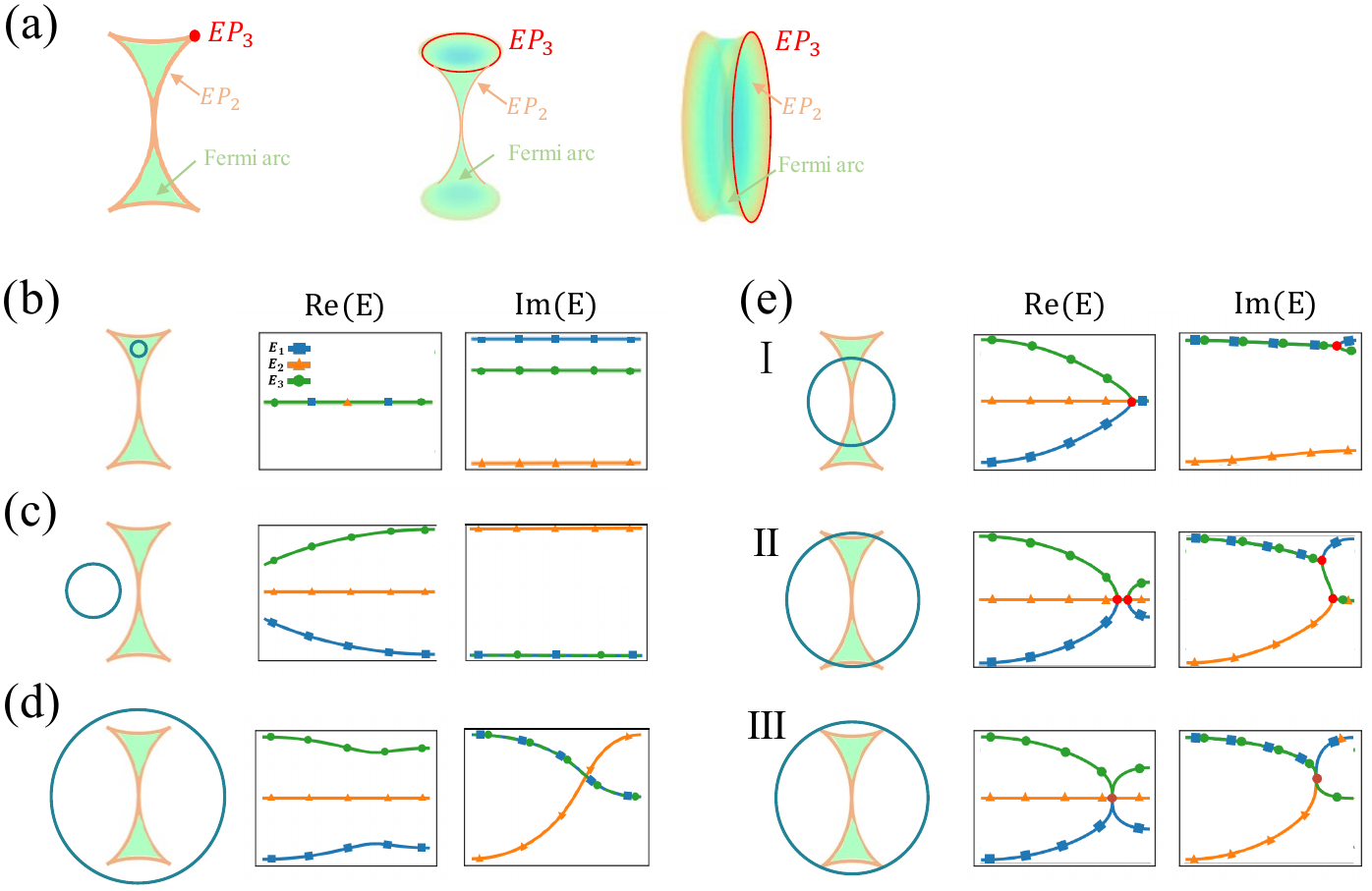}
		\caption{(a) Two-dimensional and three-dimensional projection diagram of the ES in the four-dimensional parameter space. (b) (c) The parameter manifold is enclosed by or separated from the ES, with the DD invariant being 0. (d) The parameter manifold encloses the ES and the DD invariant is 1. (e) The cases under which the topological invariant cannot be well defined, including once doubly equivalent, twice doubly equivalent, and once triply equivalent.}
		\label{FigS1}
\end{figure*}
	The topology of the Dirac monopole can be characterized by the first Chern number $C_1$, which is expressed in the two-fold integration of the Berry curvature over the surface $S^2$ in a two-dimensional system with the three-dimensional Hilbert space, namely, 
	\begin{eqnarray}
		C_1=\frac{1}{2\pi}\int_{S^2}\mathcal{F}_{\mu\nu}\textrm{d}q^\mu\wedge \textrm{d}q^\nu.
	\end{eqnarray}
	While for a tensor monopole, which appears in a three-dimensional system with the four-dimensional Hilbert space, its topology can be characterized by the Dixmier-Douady (DD) invariant determined by the three-form Berry curvature:
	\begin{eqnarray}
		\mathcal{DD}=\frac{1}{2\pi^2}\int_{S^3}M_{\mu\nu\lambda}\textrm{d}q^\mu\wedge \textrm{d}q^\nu\wedge \textrm{d}q^\lambda.
	\end{eqnarray}
	The three-form curvature tensor is related to the quantum metric
	\begin{equation}
		M_{\mu\nu\lambda}=\epsilon_{\mu\nu\lambda}\left[4\sqrt{\textrm{det}(g_{\mu\nu\lambda})}\right],
	\end{equation}
	where $\epsilon_{\mu\nu\lambda}$ is the Levi-Civita symbol, whose components can be arranged into a $3\times3\times3$ array according to the order between $\mu,\nu,\lambda$. The quantum metric in the three-dimensional system is written as a $3\times3$ form:
	\begin{eqnarray}\label{s4}
		g_{\mu\nu\lambda}=\begin{pmatrix}
			g_{\mu\mu}&g_{\mu\nu}&g_{\mu\lambda}\\
			g_{\nu\mu}&g_{\nu\nu}&g_{\nu\lambda}\\
			g_{\lambda\mu}&g_{\lambda\nu}&g_{\lambda\lambda}
		\end{pmatrix},
	\end{eqnarray}
	where $g_{jk}$ $(j,k= \mu, \nu, \lambda)$ is the conventional quantum metric tensor, which defines the distance between two nearby states $|\psi_n\rangle$ and $|\psi_{n+1}\rangle$.
	
	There exists another approach to calculate the curvature tensor by the two-form Berry curvature $\mathcal{F}_{jk}$ ($j,k = \mu, \nu, \lambda$),
	\begin{eqnarray}
		M_{\mu\nu\lambda}=-\frac{1}{2}\left(\mathcal{F}_{\mu\nu}+\mathcal{F}_{\lambda\mu}\right).
		\label{eq05}
	\end{eqnarray}
	The elements in the matrix of (\ref{s4}), $g_{jk}$ $(j,k= \mu, \nu, \lambda)$, correspond to the real part of the quantum geometric tensor $\chi_{\mu\nu}$, whose imaginary part represents the two-form Berry curvature $\mathcal{F}_{jk}$, i.e., 
	\begin{eqnarray}
		\chi_{\mu\nu}=g_{\mu\nu}+i\mathcal{F}_{\mu\nu}/2.
	\end{eqnarray}
	In the Hermitian systems, $g_{jk}$ and $\mathcal{F}_{jk}$ are defined as
	\begin{eqnarray}
		&&g_{\mu\mu}=\textrm{Re}\left[\sum_{n\neq-}\frac{|\langle \psi_{-}|\partial_\mu H|\psi_n\rangle|^2}{(E_{-}-E_n)^2}\right],\nonumber\\
		\nonumber\\
		&&g_{\mu\nu}=\textrm{Re}\left[\sum_{n\neq-}\frac{|\langle \psi_{-}|\partial_\mu H|\psi_n\rangle\langle \psi_{-}|\partial_\nu H|\psi_n\rangle|}{(E_{-}-E_n)^2}\right],\nonumber\\
		&&\mathcal{F}_{\mu\nu}=\textrm{Im}\left[\sum_{n\neq-}\frac{|\langle \psi_{-}|\partial_\mu H|\psi_n\rangle\langle \psi_{-}|\partial_\nu H|\psi_n\rangle|}{(E_{-}-E_n)^2}\right].
	\end{eqnarray}
	For the NH systems, due to the difference between left and right eigenvectors, the definitions are modified as
	\begin{eqnarray}
		&&g_{\mu\mu}=\textrm{Re}\left[\sum_{n\neq-}\frac{|\langle \psi_{-}^{L}|\partial_\mu H|\psi_n\rangle|^2}{(E^L_{-}-E_n)(E_{-}-E_n^L)}\right],\nonumber\\
		\nonumber\\
		&&g_{\mu\nu}=\textrm{Re}\left[\sum_{n\neq-}\frac{|\langle \psi_{-}^{L}|\partial_\mu H|\psi_n\rangle||\langle \psi_{-}^{L}|\partial_\nu H|\psi_n\rangle|}{(E^L_{-}-E_n)(E_{-}-E_n^L)}\right],\nonumber\\
		&&\mathcal{F}_{\mu\nu}=\textrm{Im}\left[\sum_{n\neq-}\frac{|\langle \psi_{-}^{L}|\partial_\mu H|\psi_n\rangle||\langle \psi_{-}^{L}|\partial_\nu H|\psi_n\rangle|}{(E^L_{-}-E_n)(E_{-}-E_n^L)}\right].
		\label{eq9}
	\end{eqnarray}
	Here $\langle\psi_n^L|$ denotes the normalized left eigenvector of $|\psi_n\rangle$, which satisfies $\langle\psi_n^L|\psi_m\rangle=\delta_{mn}$, and $E_n^L$ is the eigenenergy of $\langle\psi_n^L|$ satisfying $\langle\psi_n^L|H=\langle\psi_n^L|E_n^L$. 
	
	A three-band NH Hamiltonian can be constructed by using the Gell-Mann matrices:
    \begin{eqnarray}
		&&\lambda_1=\begin{pmatrix}
			0&1 &0 \\
			1&0&0\\
			0&0&0
		\end{pmatrix},
        \lambda_2=\begin{pmatrix}
			0&-i &0 \\
			i&0&0\\
			0&0&0
		\end{pmatrix},
        \lambda_3=\begin{pmatrix}
			1&0 &0 \\
			0&-1&0\\
			0&0&0
		\end{pmatrix},\nonumber\\
        &&\lambda_4=\begin{pmatrix}
			0&0 &1 \\
			0&0&0\\
			1&0&0
		\end{pmatrix},
        \lambda_5=\begin{pmatrix}
			0&0 &-i \\
			0&0&0\\
			i&0&0
		\end{pmatrix},
        \lambda_6=\begin{pmatrix}
			0&0 &0 \\
			0&0&1\\
			0&1&0
		\end{pmatrix},\nonumber\\
        &&\lambda_7=\begin{pmatrix}
			0&0 &0 \\
			0&0&-i\\
			0&i&0
		\end{pmatrix},
        \lambda_8=\begin{pmatrix}
			\frac{1}{\sqrt{3}}&0 &0 \\
			0&\frac{1}{\sqrt{3}}&0\\
			0&0&-\frac{2}{\sqrt{3}}
		\end{pmatrix},
	\end{eqnarray}
    and written as $H = \vec{q}\cdot \vec{\lambda}+i\kappa\lambda_8$. 
	Due to the U(1) symmetry of the system in the parameter spaces of \{$\vec{q_1}$, $\vec{q_2}$\} and \{$\vec{q_3}$, $\vec{q_4}$\}, without loss of generality, the NH Hamiltonian in the four-dimensional parameter space can be written as 
    \begin{eqnarray}
        H/\hbar=\begin{pmatrix}
			i\kappa/\sqrt{3}&\Omega_1 &0 \\
			\Omega_1&i\kappa/\sqrt{3}&\Omega_2\\
			0&\Omega_2&-2i\kappa/\sqrt{3}
		\end{pmatrix}.
        \label{S9}
    \end{eqnarray}
	In \{$\vec{q_1}$, $\vec{q_3}$\}, due to the introduction of the NH term, the tensor monopole expands to the four vertices in the two-dimensional parameter space of \{$\vec{q_1}$, $\vec{q_3}$\}, which is projected from the ES of EP3s in the four-dimensional parameter space of \{$\vec{q_1}$, $\vec{q_2}$, $\vec{q_3}$, $\vec{q_4}$\}, as shown in Fig.~\ref{FigS1} (a). The vertices are connected by a closed curve of EP2s and the Fermi arc exists inside the closed curve.

	For the peculiar eigenenergy structures, the four-dimensional hypersphere parameter space we constructed can be divided into the following three cases (only the projection of the four-dimensional parameter space on \{$\vec{q_1}$, $\vec{q_3}$\} are drawn and shown in the plots).
	
	(1) When the parameter manifold is located inside the Fermi arc or separated from the ES, as shown in Fig.~\ref{FigS1} (b) and (c), the corresponding DD invariant is 0, and the system is topologically trivial.
	
	(2) When the parameter manifold completely encloses the ES, as shown in Fig.~\ref{FigS1} (d), the DD invariant is 1, and the system is topologically non-trivial.
	
	(3) When the parameter manifold intersects the exceptional surface, as shown in Fig.~\ref{FigS1} (e), according to the exceptional features of the intersection point, the eigenenergies undergo two-fold or three-fold degeneracy, for which the involved cases are once doubly equivalent, twice doubly equivalent, and once triply equivalent. At this point, the distribution of eigenenergies before and after the intersection cannot be clearly distinguished, thus the DD invariant and thus the relevant topology of this system cannot be well defined.
	
	\subsection{The Berry phase of the Möbius-like eigenenergies}
    In this part, we present another type of the characterization for the topology, namely the Berry phase, which distinguishes from the DD invariant, which can be manifested in the peculiar eigenenergy structures in the ES: the Möbius-like band. Firstly, we study a two-level Hamiltonian of  
    \begin{figure}[b!] 
		\centering
		\includegraphics[width=3.2in]{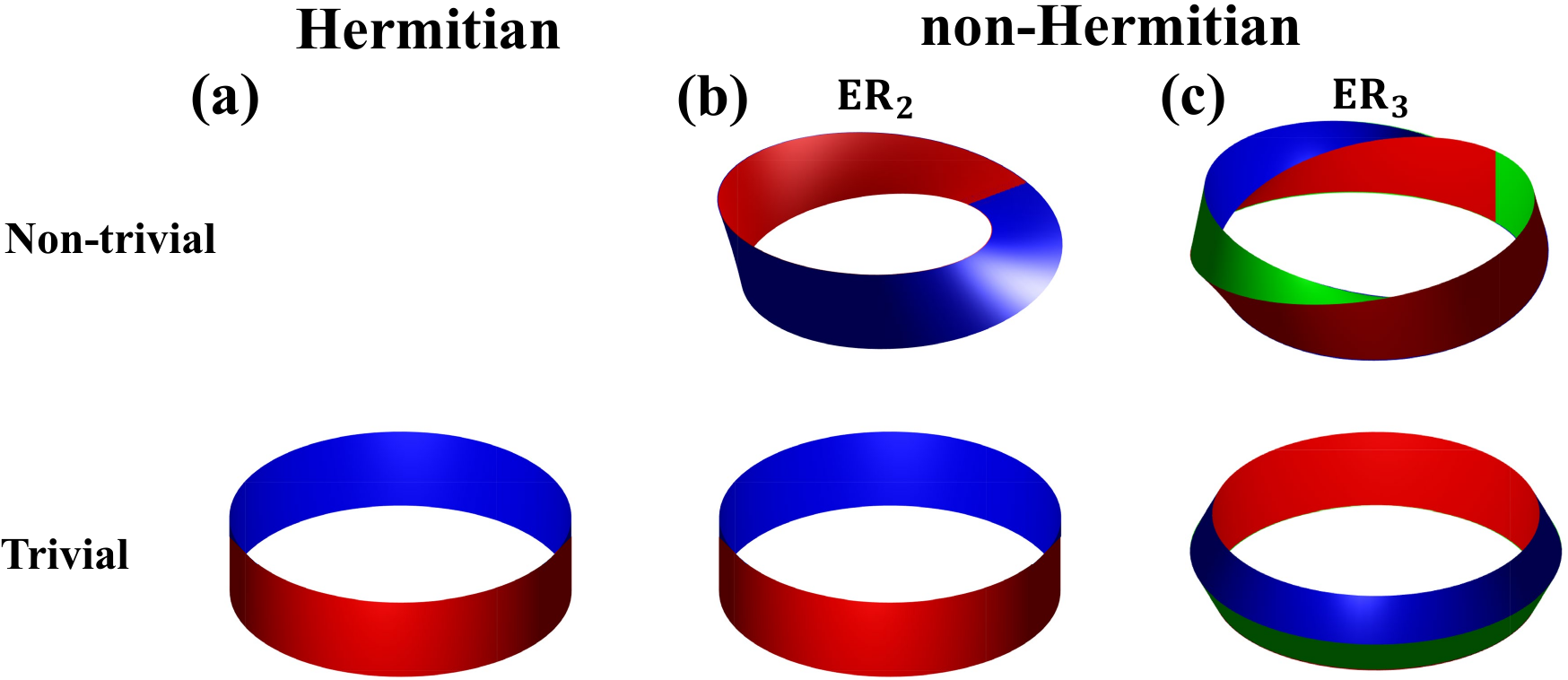}
		\caption{Schematic diagram of the eigenenergies distribution, corresponding to different cases of (a) the Hermitian system, (b) an ER in the three-dimensional system and (c) an ES in the four-dimensional system. The three different colors represent the three eigenenergy bands. The first and second rows correspond to the cases where the topology is non-trivial and trivial, respectively.}
		\label{FigS2}
	\end{figure}
	\begin{eqnarray}
H_2/\hbar=R\left(\textrm{sin}\theta\textrm{cos}\phi\hat{\sigma}_x+\textrm{sin}\theta\textrm{sin}\phi\hat{\sigma}_y+\textrm{cos}\theta\hat{\sigma}_z\right),
        \label{eq_h2}
	\end{eqnarray}
    \begin{figure*}[t!] 
		\centering
		\includegraphics[width=6.2in]{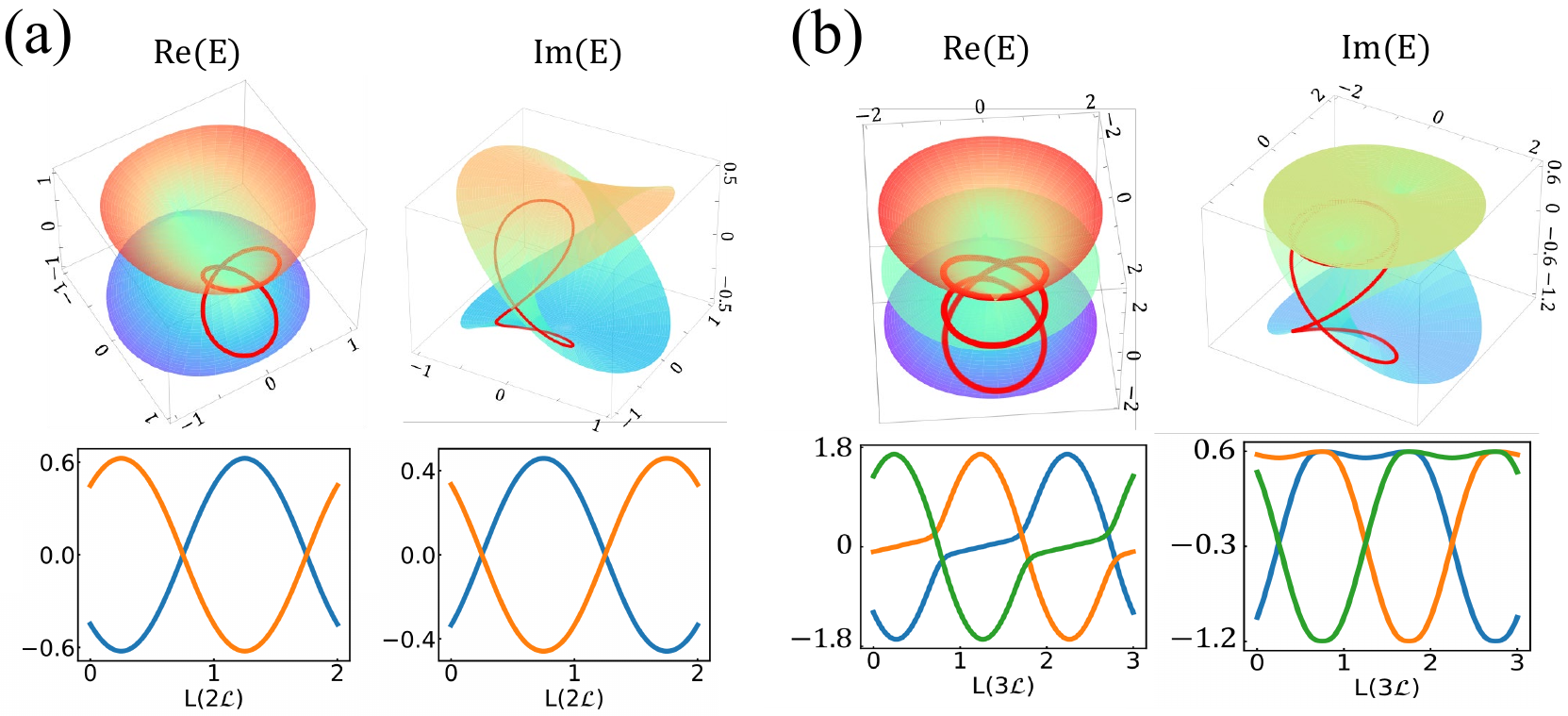}
		\caption{The real and imaginary parts of eigenenergies in the Riemann surface of (a) the 2D ER and (b) the 3D ES.}
		\label{FigS3}
    \end{figure*}
where $R$ is the amplitude of the Bloch vector to locate on the ($\theta$, $\phi$) state of the Bloch sphere, $\hat{\sigma}_{j}$ ($j = x, y, z$) are the Pauli operators.   
The degeneracy of the Hamiltonian (\ref{eq_h2}) is regarded as a Dirac monopole. The eigenenergy structures are shown in Fig.~\ref{FigS2} (a). The Berry phase in such a case is written as
    \begin{eqnarray}
        \mathcal{P} &=& -\textrm{Im}\textrm{ln}[\langle \psi_0|\psi_1\rangle\cdot\cdot\cdot\langle \psi_n|\psi_0\rangle]\nonumber\\
			&=&-\textrm{Im}\sum_{i=0}^{n}\textrm{ln}\langle\psi_i|\psi_{i+1}\rangle\nonumber\\
            &=&-\oint i\langle\psi_n|\partial_\theta\psi_n\rangle\textrm{d}\theta,
    \end{eqnarray}
    which is always 0. However, in the two-level NH Hamiltonian
    \begin{eqnarray}
        H_{2}^N=H_2+i\hbar\gamma\sigma_z,
    \end{eqnarray}
    the degeneracy as the Dirac monopole expands to an exceptional ring (ER) in \{$\left\langle\vec{\sigma_x}\right\rangle$, $\left\langle\vec{\sigma_y}\right\rangle$\}. If the parameter loop in the three-dimensional space of \{$\left\langle\vec{\sigma_x}\right\rangle$, $\left\langle\vec{\sigma_y}\right\rangle$, $\left\langle\vec{\sigma_z}\right\rangle$\} is nested in the ER, the eigenenergies undergo a flip like the Möbius band after one loop, as shown in Fig.~\ref{FigS2} (b). The eigenvectors come back to the origin after a whole evolution with two loops ($2\mathcal{L}$), as depicted in Fig.~\ref{FigS3} (a), accumulating a Berry phase
    \begin{eqnarray}
        \mathcal{P}=\oint i\langle\psi_n^L|\partial_\theta\psi_n\rangle\textrm{d}\theta
    \end{eqnarray}
    of $\pi$. The evolution of the eigenenergies and the related topological properties exhibits more unique features, as compared to the Hermitian cases. Whereas if the parameter loop is not nested in the ER, the Berry phase keeps 0 and the system is trivial.

    Furthermore, when described with a NH three-dimensional Hamiltonian of Eq. (\ref{S9}), the tensor monopole appears in the form of an exceptional surface, located in the the four-dimensional parameter space of \{$\vec{q_1}$, $\vec{q_2}$, $\vec{q_3}$, $\vec{q_4}$\}. If the parameter loop is constructed in the three-dimensional space \{$\vec{q_1}$, $\vec{q_2}$, $\vec{q_8}$\} nested with the two-dimensional projection ring of the ES in \{$\vec{q_1}$,$\vec{q_2}$\}, the eigenenergies of the system undergo a Möbius-like flip after a loop (Fig.~\ref{FigS2} (c)), and turn back to their origins after three loops, accumulating a Berry phase of $2\pi$, as illustrated in Fig.~\ref{FigS3} (b).

    Clearly, the two topological variants, namely the DD variant and the Berry phase, respectively characterize two unique topological properties of the ES in NH systems.

	\section{The protocol for the experimental implementation }
    In this part, we describe how the topological properties of the ES can be experimentally characterized. Here we consider a superconducting circuit architecture where a frequency-tunable Xmon qubit ($Q$), whose sweet frequency point is around $6.0$ GHz, is coupled to a bus resonator ($R_1$) with frequency of $5.58$ GHz and \textcolor{red}{to} a readout resonator ($R_2$) with  frequency around $6.66 \thicksim 6.85$ GHz \cite{PhysRevLett.131.260201,han2024measuring}. The interaction Hamiltonian depicting the coupling of $Q$ to $R_1$ and $R_2$ can be modeled as     
    \begin{eqnarray}
		H_I/\hbar= \sum_{j=1,2}g_j e^{i\Delta_j t} \hat{a}_j^\dagger \hat{\sigma}^- + H.c.
		\label{B1}
	\end{eqnarray}   
    In (\ref{B1}), $\hat{a}_j^\dagger$ ($\hat{a}_j$) is the creation (annihilation) operator for the photon of resonator $j$, $\hat{\sigma}^-$ ($\hat{\sigma}^+$) denotes the lowing (raising) operator for qubit, $g_1$ ($g_2$) is the coupling strength between $Q$ and $R_1$ ($R_2$) and typically has the value of about $2\pi \times 20$ ($2\pi \times 40$) MHz in the experiments we implemented \cite{PhysRevLett.131.260201,han2024measuring}, and $\Delta_j = \omega_q - \omega_{r_j}$ represents the detuning between the frequencies of $Q$ ($\omega_e/2\pi$) and $R_j$ ($\omega_{r_j}/2\pi$).  
    
\begin{figure}[b!] 
		\centering
		\includegraphics[width=3.2in]{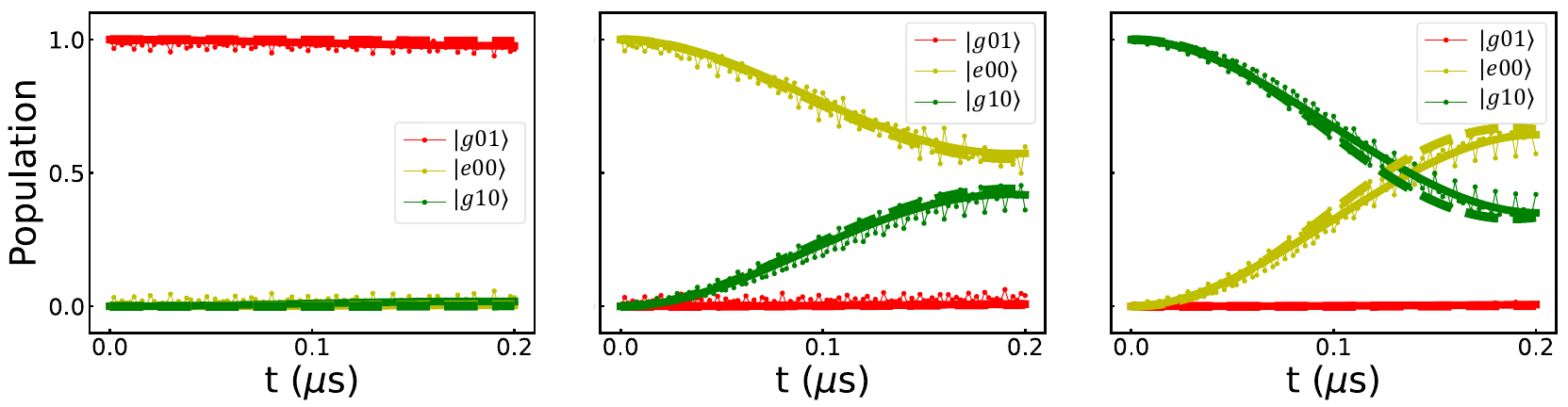}
		\caption{Population distribution of $|e00\rangle$, $|g10\rangle$ and $|g01\rangle$ for different initial states at time-dependent evolution. The oscillating-solid, dashed- and dotted-lines represent the outcomes with the full Hamiltonian, the ideal situation and the fitted case, respectively.}
		\label{FigS4}
\end{figure}

In order to establish arbitrary parameter manifolds or loops, two cosines pulses are applied to the z-control of $Q$ \cite{PhysRevLett.131.260201,han2024measuring}, resulting in a modulation to the frequency of $Q$ in the form of 
    	\begin{eqnarray}
		\omega_e\left(t\right)=\omega_q+\varepsilon_1\textrm{cos}
		\left(\nu_1 t\right)+\varepsilon_2\textrm{cos}\left(\nu_2 t\right),
		\label{S13}
	\end{eqnarray}
	where $\varepsilon_{1,2}$ and $\nu_{1,2}$ denote the modulation amplitudes and frequencies, respectively. In such a case, the interaction Hamiltonian (\ref{B1}) is reduced to
        \begin{eqnarray}
		H_I/\hbar&=&\sum_{m,n}^{1,2}g_m\hat{a}_m^\dagger\hat{\sigma}^-\sum_{k}J_k\left(\mu_n\right)\textrm{e}^{-i\left(k\nu_n-\Delta_m\right)t}+H.c.,\nonumber\\
	\end{eqnarray}
	where $\mu_n=\varepsilon_n/\nu_n$, $\Delta_m=\omega_{r_m}-\omega_{q}$, and $J_k(\mu_n)$ denotes the n-th Bessel function of the first kind for the Jacobi-Anger expansion obtained from the formula $\textrm{e}^{i\mu\textrm{sin}x}=\sum_{k = -\infty}^{\infty}J_k\left(\mu\right)\textrm{e}^{ikx}$. 
	We may set the modulation frequencies to satisfy $\nu_1=\Delta_1/2$ and $\nu_2=-\Delta_2$. In such a way, all the high-frequency oscillation terms can be ignored, thus the interaction Hamiltonian describing the unitary dynamics is simplified as 
    \begin{equation}
    H_I/\hbar=\sum_{j=1,2}\Omega_j\hat{a}_j^\dagger\hat{\sigma}^- + H.c., 
    \end{equation}
    where the availably modulated coupling coefficient $\Omega_j$ of the 1st-order lower amplitudes as compared to $g_j$ can be reached according to our previous experiments \cite{PhysRevLett.131.260201,han2024measuring}. We take into account the energy relaxation of $R_2$, which is of the same order of $\Omega_j$ (a few MHz) and predominates the other channels of decoherence (including energy relaxation and dephasing of both $Q$ and $R_1$, dephasing of $R_2$), the rates for which are on the order of one-percent or one-tenth of the maxima of $\Omega_j$ \cite{PhysRevLett.131.260201,han2024measuring}. Under the circumstance, the dynamics of the whole system is approximately modeled by the Lindblad master equation:
    \begin{equation}
    \frac{d\rho(t)}{dt} = -\frac{i}{\hbar}[H_I^{\prime},\rho(t)] + \kappa_{d,2} \hat{a}_2 \rho(t)  \hat{a}_2^{\dagger},
     \end{equation}
    where 
    \begin{equation}
    H_I^{\prime} = H_I - i\hbar\frac{\kappa_{d,2}}{2}\hat{a}_2^{\dagger}\hat{a}_2,
    \label{B6}
    \end{equation}
    with $\kappa_{d,2}$ being the energy relaxation rate for $R_2$. The non-Hermiticity and inherent exceptional features dominate only when no-jump trajectories take place during the whole dissipative dynamics, under which condition the `tailored dissipative dynamics' functions within the state subspace: $S \equiv \{|e00\rangle,|g10\rangle,|g01\rangle\}$, controlled by the conditional Hamiltonian of (\ref{B6}). For the experimental implementation, the case of the no-jump trajectory can exactly be identified by the method of postselection \cite{PhysRevLett.131.260201}. Namely this is achieved by extracting the information of the complete state of the system and discarding the components of $|g00\rangle$, and finally renormalizing states of the system within the subspace of $S$. Thus at an arbitrary time, the state of the system, which starts from a designated initial state and evolves following the NH dynamics controlled by the conditional Hamiltonian of (\ref{B6}), can be reconstructed. As an arbitrary initial state can be expanded with a specific linear combination of right eigenvectors, identifying the states of the system at different times should provide sufficient information for extracting both right eigenvectors and eigenenergies of the system. In addition, left eigenvectors can also be obtained by resorting to the biorthogonal condition \cite{moiseyev_non-hermitian_2011}. This means that, the eigenvectors and the corresponding evolution eigenenergies at each point in the parameter space can be acquired. Though in practice this is potentially limited by the implemented parameters.
       
	\begin{figure}[t!] 
		\centering
		\includegraphics[width=3.2in]{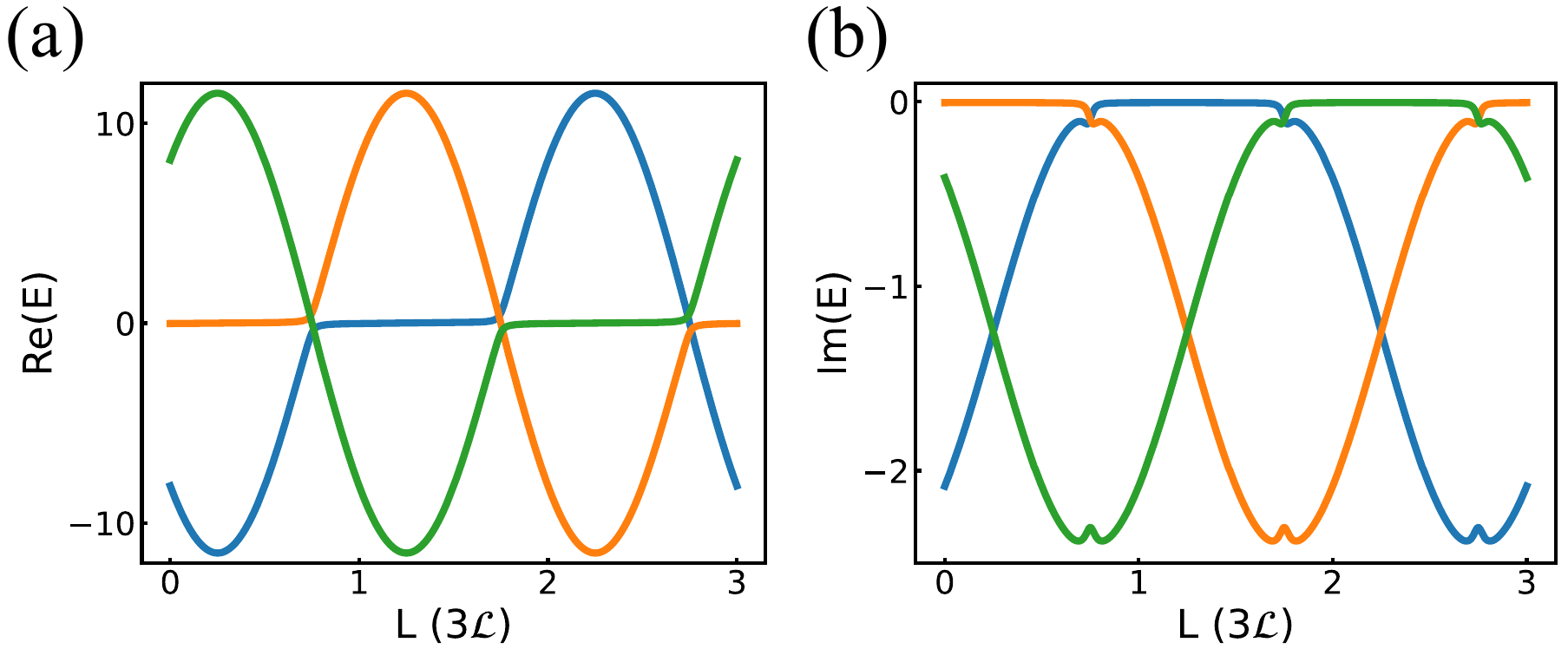}
		\caption{The fitted real (a) and imaginary (b) parts of the three eigenenergies along the whole evolution path of three loops $3\mathcal{L}$. }
		\label{FigS5}
	\end{figure}
\begin{figure}[b!] 
	\centering
	\includegraphics[width=3.2in]{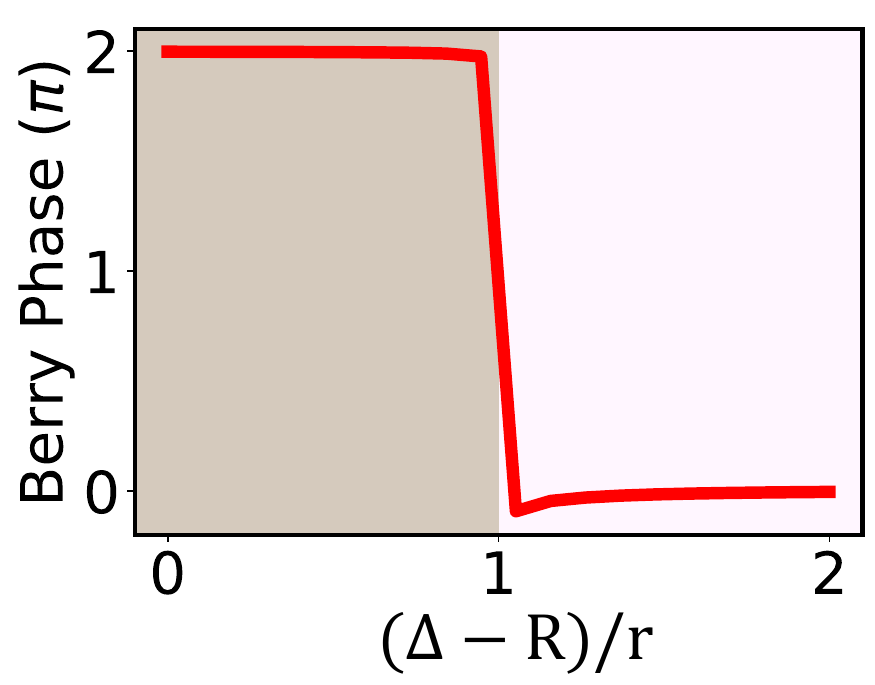}
	\caption{The calculated and fitted Berry phase as $\Delta$ increases. When $(\Delta-R)/r<1$, the parameter path encircles the ER and the Berry phase is $2\pi$; while $(\Delta-R)/r>1$, the Berry phase is 0.}
	\label{FigS6}
\end{figure}
    Take the Berry phase for an analysis. The control for establishing the loop satisfies the condition of $[g_1J_2(\mu_1)=|\Omega_1|$, $g_2J_{-1}(\mu_2)=R\textrm{sin}\theta+\Delta$ and $q_{\perp} = R\textrm{cos}\theta]$, with the set of experimentally available parameters chosen as follows: $\omega_{r_1}=2\pi\times5.58$ GHz, $\omega_{r_2}=2\pi\times6.66$ GHz, $g_1=2\pi\times20$ MHz, $g_2=2\pi\times40$ MHz, $\kappa_{d,2}=5$ MHz, $\omega_{q}=2\pi\times5.86$ GHz, $\nu_1=2\pi\times0.4$ GHz and $\nu_2=2\pi\times0.28$ GHz. The temporal evolution of the system state is determined by the conditional NH dynamics dominated by (\ref{B6}). Starting respectively from the three initial states $|e00\rangle$, $|g10\rangle$ and $|g01\rangle$, the three corresponding states at designated times can be completely obtained through the method of quantum state tomography \cite{PhysRevLett.131.260201}. The instance of these evolutions in the case of $\theta = 0$ set by the control of system parameters at $t=0.2$ $\mu$s are depicted in Fig~\ref{FigS4}, where the oscillating-solid-, dashed- and dotted-lines correspond to the outcomes with the full Hamiltonian, the ideal situation and the fitted case, respectively. We assume the system parameters are controlled in such a way that $\theta$ changes smoothly from 0 to $6\pi$. For a specific value of $\theta$, the real and imaginary parts of the eigenenergies can be extracted and are shown in Fig~\ref{FigS5}, exhibiting a Möbius-like structure. In this case, the Berry phase calculated based on the extracted results is $0.998\times2\pi$, with the slight deviation from unity due to the imperfection of the dynamics process. As $\Delta$ is controlled to gradually increase, the parameter loop no longer encircles the ER, in which case the system becomes trivial and the thus-obtained Berry phase is zero. The criticality at which the transition of the Berry phase happens is shown in Fig~\ref{FigS6}.
    
    \bibliography{reference}